\documentclass[final,twocolumn]{elsarticle}

\usepackage{amssymb}
\usepackage{graphicx}
\usepackage{dcolumn}
\usepackage{bm}
\usepackage{gensymb}
\usepackage{lineno}

\begin{document}

\begin{frontmatter}



\title{Measurement of spin-flip probabilities for ultracold neutrons
  interacting with nickel phosphorus coated surfaces}


\author[LANL]{Z.~Tang}
\author[IU]{E.~R.~Adamek}
\author[NCSU]{A.~Brandt}
\author[IU]{N.~B.~Callahan}
\author[LANL]{S.~M.~Clayton}
\author[LANL]{S.~A.~Currie}
\author[LANL]{T.~M.~Ito\corref{cor1}}
\ead{ito@lanl.gov}
\author[LANL]{M.~Makela}
\author[KEK]{Y.~Masuda}
\author[LANL]{C.~L.~Morris}
\author[LANL]{R.~Pattie Jr.}
\author[LANL]{J.~C.~Ramsey}
\author[LANL,IU]{D.~J.~Salvat}
\author[LANL]{A.~Saunders}
\author[NCSU]{A.~R.~Young}

\address[LANL]{Los Alamos National Laboratory, Los Alamos, New Mexico 87545,
 USA}
\address[IU]{Indiana University, Bloomington, Indiana 47405, USA}
\address[NCSU]{North Carolina State University, Raleigh, North Carolina,
27695, USA}
\address[KEK]{High Energy Accelerator Research Organization, Tsukuba, Ibaraki 305-0801, Japan}
\cortext[cor1]{Corresponding author}

\begin{abstract}
We report a measurement of the spin-flip probabilities for ultracold
neutrons interacting with surfaces coated with nickel phosphorus. For
50~$\mu$m thick nickel phosphorus coated on stainless steel, the
spin-flip probability per bounce was found to be $\beta_{\rm
  NiP\;on\;SS} = (3.3^{+1.8}_{-5.6}) \times 10^{-6}$. For 50~$\mu$m
thick nickel phosphorus coated on aluminum, the spin-flip probability
per bounce was found to be $\beta_{\rm NiP\;on\;Al} =
(3.6^{+2.1}_{-5.9}) \times 10^{-6}$. For the copper guide used as
reference, the spin flip probability per bounce was found to be
$\beta_{\rm Cu} = (6.7^{+5.0}_{-2.5}) \times 10^{-6}$.  The results on
the nickel phosphorus-coated surfaces may be interpreted as upper
limits, yielding $\beta_{\rm NiP\;on\;SS} < 6.2 \times 10^{-6}$ (90\%
C.L.) and $\beta_{\rm NiP\;on\;Al} < 7.0 \times 10^{-6}$ (90\% C.L.)
for 50~$\mu$m thick nickel phosphorus coated on stainless steel and
50~$\mu$m thick nickel phosphorus coated on aluminum, respectively.
Nickel phosphorus coated stainless steel or aluminum provides a
solution when low-cost, mechanically robust, and non-depolarizing UCN
guides with a high-Fermi-potential are needed.
\end{abstract}

\begin{keyword}



\end{keyword}

\end{frontmatter}


\section{\label{sec:intro}Introduction}
Ultracold neutrons (UCNs) are defined operationally to be neutrons of
sufficiently low kinetic energies that they can be confined in a
material bottle, corresponding to kinetic energies below about
340~neV. UCNs are playing increasingly important roles in the studies
of fundamental physical interactions (for recent reviews, see {\it
  e.g.}  Refs.~\cite{DUB11,YOU14}).

Experiments using UCNs are being performed at UCN facilities around
the world, including Institut Laue-Langevin (ILL)~\cite{STY86}, Los
Alamos National Laboratory (LANL)~\cite{SAU13}, Research Center for
Nuclear Physics (RCNP) at Osaka University~\cite{MAS12}, Paul Scherrer
Institut (PSI)~\cite{BEC15}, and University of Mainz~\cite{LAU13}. One
important component for experiments at such facilities is the UCN
transport guides. These guides are used to transport UCNs from a
source to experiments and from one part of an experiment to
another. For applications that require spin polarized UCNs, it is
important that UCNs retain their polarization as they are transported
(see {\it e.g.} Ref.~\cite{SAL14}).

In other applications in which polarized UCNs are stored for extended
periods of time, such as neutron electric dipole moment experiments
(see {\it e.g.} Ref.~\cite{BAK06}) and the UCNA experiment (see {\it
  e.g.}  Ref.~\cite{PLA12}), it is also important that UCNs remain
highly polarized while they are stored in a material bottle. In some
cases, the depolarization of UCNs due to wall collisions is one of the
dominant sources of systematic uncertainties in the final result of
the experiment~\cite{MEN13}.

Because of its importance for these applications, the study of
spin-flip probabilities and the possible mechanisms for spin flip is
an active area of research~\cite{SER00,POK02,SER03,ATC07}. The
spin-flip probability per bounce has been measured for various
materials. References~\cite{SER00,SER03} report results on beryllium,
quartz, beryllium oxide, glass, graphite, brass, copper, and Teflon,
whereas Ref.~\cite{ATC07} discusses results on diamond-like carbon
(DLC) coated on aluminum foil and on polyethylene terephthalate (PET)
foil. In addition, measurements of the spin-flip probability per
bounce have been performed for DLC-coated quartz~\cite{MAK05},
stainless steel, electropolished copper, and DLC-coated
copper~\cite{RIO09}. (The results from Ref.~\cite{RIO09} are available
in Ref.~\cite{HOL12}.) The reported values for the spin-flip
probability per bounce are on the order of $10^{-6}-10^{-5}$ for all
of these materials with the exception of stainless steel, for which a
reported preliminary value for the spin-flip probability per bounce is
on the order of $10^{-3}$~\cite{RIO09}, two to three orders of
magnitude larger.

The spin-flipping elastic or quasielastic incoherent scattering from
protons in surface hydrogen contamination has been considered to be a
possible mechanism for UCN spin flip upon interaction with a
surface~\cite{SER00,POK02,SER03,ATC07}. So far, however, data and
model calculations have not been in agreement.

Another possible mechanism is Majorana spin flip~\cite{VLA61} due to
magnetic field inhomogeneity near material surfaces, from
ferromagnetic impurities or magnetization of the material itself. In
addition, the sudden change in direction that occurs when a UCN
reflects from a surface can cause a spin flip in the presence of
moderate gradients~\cite{POK02}. Gamblin and Carver~\cite{GAM65}
discuss such an effect for $^3$He atoms. The high spin-flip
probability observed for stainless steel in a large holding
field~\cite{RIO09} is likely due to magnetic field inhomogeneity near
material surfaces or magnetization of the material itself.

Recently, based on the suggestion from Ref.~\cite{MAS13}, we have
identified nickel phosphorus (NiP) coating to be a promising UCN
coating material with a small loss per bounce and a high Fermi
potential~\cite{PAT15}. The Fermi potentials of NiP samples with a
phosphorus content of 10.5 wt.\% (18.2 at.\%) coated on
stainless steel and on aluminum were both found to be
$\approx$213~neV~\cite{PAT15}, which is consistent with the calculated
value and is to be compared to 188~neV for stainless steel and 168~neV
for copper. NiP coating is extremely robust and is widely used in
industrial applications. It is highly attractive from a practical
point of view because the coating can be applied commercially in a rather
straightforward chemical process, and there are numerous vendors that
can provide such a service economically. It is important that the
coating process not use neutron absorbing materials such as
cadmium. Furthermore, alloying nickel with phosphorus lowers its Curie
temperature (see {\it e.g.}  Refs.~\cite{ALB67,BER78,HUM98}). As a
result, when made with high enough phosphorus content, NiP is known to
be non-magnetic at room temperature. Therefore, it is of great
interest to study its UCN spin depolarization properties and how
depolarization depends on the material used for the substrate.

In this paper, we report a measurement of the spin-flip probabilities
for UCNs interacting with NiP-coated surfaces. We investigated
aluminum and stainless steel as the substrate. The results obtained
with NiP-coated surfaces are compared to those obtained with copper
guides. This measurement was performed as part of development work for
a new neutron electric dipole moment experiment at the LANL UCN
facility~\cite{ITO14} and an associated UCN source
upgrade~\cite{ITO15}.

This paper is organized as follows. In Sec.~\ref{sec:experiment}, the
experimental apparatus and method are
described. Section~\ref{sec:analysis} describes the analysis of the
data. In~Sec.~\ref{sec:discussion}, we discuss the implication of the
results on future experiments using UCNs. Section~\ref{sec:summary}
provides a short summary of the content of this paper.

\section{\label{sec:experiment}Experiment}
\subsection{Apparatus and method}
The measurement was performed at the LANL UCN
facility~\cite{SAU13}. Spallation neutrons produced by a pulsed
800-MeV proton beam striking a tungsten target were moderated by
beryllium and graphite moderators at ambient temperature and further
cooled by a cold moderator that consisted of cooled polyethylene
beads. The cold neutrons were converted to UCNs by a solid deuterium
(SD$_2$) converter. UCNs were directed upward 1~m along a vertical
guide coated with $^{58}$Ni and then 6~m along a horizontal guide made
of stainless steel before exiting the biological shield. At the bottom
of the vertical UCN guide was a butterfly valve that remained closed
when there was no proton beam pulse striking the spallation target in
order to keep the UCNs from returning to the SD$_2$ where they would
be absorbed.

\begin{figure}
\centering
\includegraphics[width=3in]{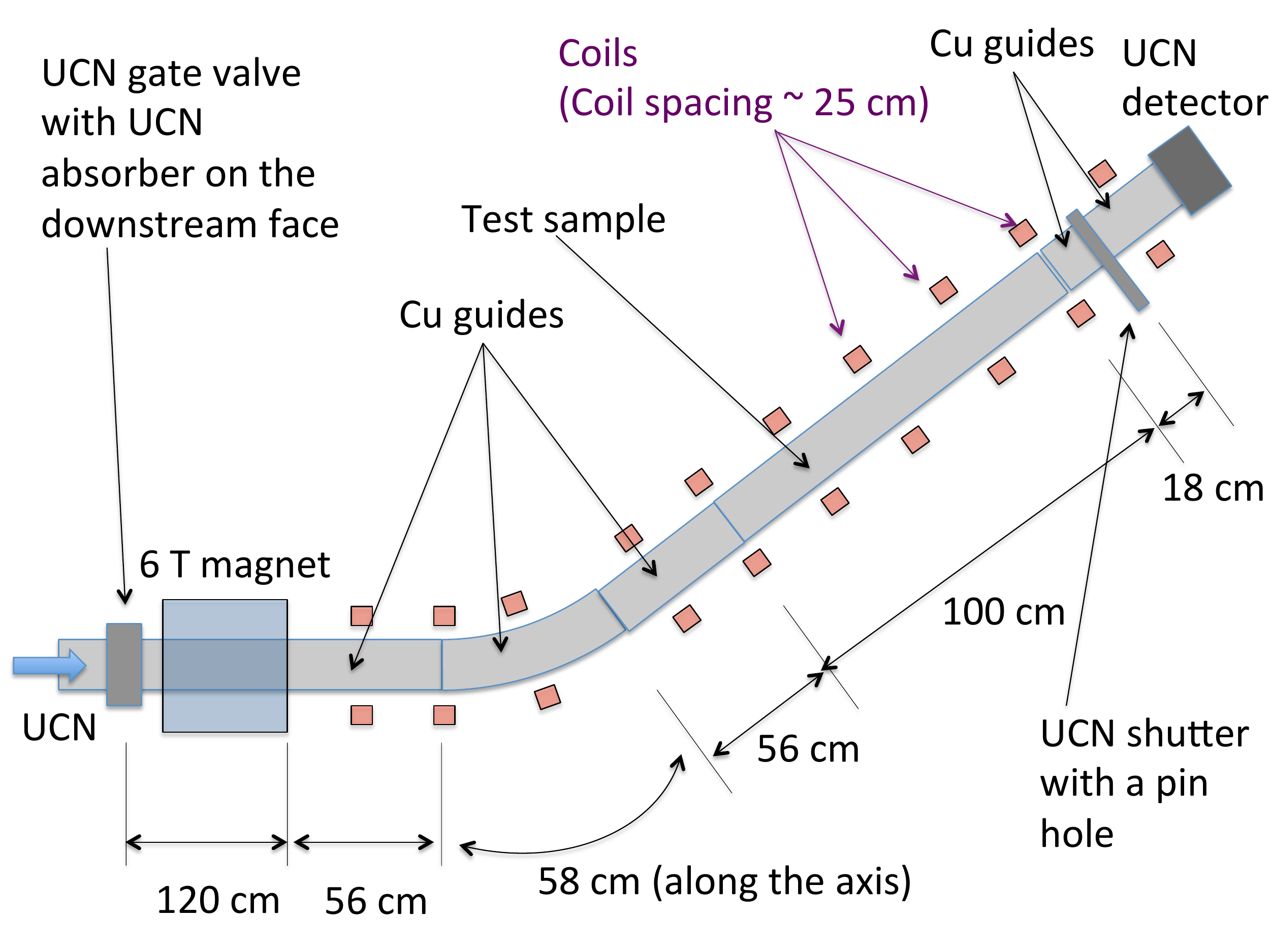}
\caption{Schematic diagram of the experimental setup\label{fig:schematic2}}
\end{figure}

A schematic diagram of the experimental setup for the depolarization
measurement is shown in Fig.~\ref{fig:schematic2}.
When the UCN gate valve was opened, UCNs transported from the source
entered the apparatus. UCNs in one spin state, the so-called
``high-field seekers'' for which $\bm{\mu}\cdot\bm{B}<0$ where
$\bm{\mu}$ is the magnetic dipole moment of the neutron and $\bm{B}$
is a magnetic field, were able to move past the 6~T magnetic field
provided by a superconducting solenoidal magnet. UCNs in the other
spin state, the so-called ``low-field seekers'', were reflected back
by the potential barrier due to $\bm{\mu}\cdot\bm{B}>0$. For
$|\bm{B}|=6$~T, $|\bm{\mu}\cdot\bm{B}|=360$~neV, much larger than the
kinetic energy of the neutrons from the LANL UCN source, which has a
cutoff at $\approx$190~neV.

On the other side of the high-field region was a UCN guide system
(76.2~mm in OD, 72.9~mm in ID) that consisted of a set of copper guide
sections, a NiP-coated guide section, a section of a copper guide with
a UCN shutter, and a short section of a copper guide followed by a UCN
detector~\cite{WAN15}. The guides were placed in a $\sim$2~mT magnetic
field provided by a set of coils in order to retain the polarization
of the UCNs. The UCN shutter, made of copper, had a pinhole 5~mm in
diameter. The gap between the UCN shutter and the end of the guide
leading to it was measured to be $\sim$0.05~mm.

The high-field seekers were able to move freely through the high
magnetic field region, colliding with the inner walls of the guide
system. The number of collisions per second for the high-field seekers
in the guide system downstream of the 6~T field region is given by
\begin{equation}
R_{\rm hfs} = \frac{1}{4}A_{\rm tot}\,\langle v_{\rm hfs}\rangle\,n_{\rm hfs},
\end{equation}
where $A_{\rm tot}$ is the total inner surface of the system that the
high field seekers interacted with in the guide system downstream of
the 6~T field region, $\langle v_{\rm hfs}\rangle$ is the average
velocity of the high-field seekers, and $n_{\rm hfs}$ is the density
of the high-field seekers. The rate of collisions for the high field
seekers was monitored during this time by detecting neutrons that
leaked through a pinhole on the shutter in the downstream end of the
test guide assembly. If UCNs are detected at a rate of $R_{\rm h}$ and
the area of the pinhole is $A_{\rm h}$, $R_{\rm hfs}$ can be inferred
to be
\begin{equation}
\label{eq:rate}
R_{\rm hfs} = \frac{A_{\rm tot}}{A_{\rm h}}R_{\rm h},
\end{equation}
assuming that the distribution of UCNs inside the system is uniform
and isotropic. We discuss the validity of this assumption in
Sec.~\ref{sec:analysis}. For our geometry, ${A_{\rm tot}}/{A_{\rm h}}
= 3.36\times 10^4$.

While the high-field seekers collided with the inner wall of the
system at the rate of $R_{\rm hfs}$, spin-flipped neutrons were
produced at a rate of
\begin{equation}
R_{\rm dep} = R_{\rm hfs}\,\beta = \frac{A_{\rm tot}}{A_{\rm h}}R_{\rm h}\,\beta, 
\end{equation}
where $\beta$ is the probability of spin flip per bounce. While in
principle $\beta$ can depend on the neutron velocity as well as the
angle of incidence, and such dependencies can give important
information on the possible mechanism of UCN depolarization upon wall
collision, we assumed $\beta$ to be independent of the velocity and
angle of incidence in our analysis. Reference~\cite{ATC07} observed no
indication of energy dependence within the accuracy of their
measurement for the depolarization per bounce on DLC.

The spin-flipped neutrons were not able to pass through the 6~T field
region and hence became ``trapped'' in the guide assembly downstream
of the 6~T field region. The number of trapped spin-flipped neutrons
in the system at time $t$, $n_{\rm dep}(t)$, satisfies the following
differential equation:
\begin{equation}
\label{eq:diff}
\frac{dn_{\rm dep}(t)}{dt} = \beta\frac{A_{\rm tot}}{A_{\rm h}}R_{\rm
  h}(t) 
- \frac{1}{\tau_{\rm dep}}n_{\rm dep}(t),
\end{equation}
where $\tau_{\rm dep}$ is the lifetime of the spin-flipped
neutrons. If $R_{\rm hfs}$ was constant, then the number of the
spin-flipped low-field seekers built up as 
$[1-\exp(-t/\tau_{\rm dep})]$. In general, $\tau_{\rm dep}$ depends on the UCN velocity
$v$. With the velocity dependence taken into account,
Eq.~(\ref{eq:diff}) becomes
\begin{equation}
\label{eq:diff2}
\frac{\partial^2 n_{\rm dep}}{\partial v\,\partial t} = \beta\frac{A_{\rm tot}}{A_{\rm h}}P(v,t)R_{\rm h}(t) 
- \frac{1}{\tau_{\rm dep}(v)}\frac{\partial n_{\rm dep}}{\partial v},
\end{equation}
where $P(v,t)$ is the normalized velocity distribution of UCNs at the
time of spin flip, and $\int_0^{v_c}\frac{\partial n_{\rm
    dep}}{\partial v}dv = n_{\rm dep}$ where $v_{\rm c}$ is the cutoff
velocity.

The UCN gate valve upstream of the 6~T region was then closed, after
which time the high-field seekers were quickly absorbed by a thin
sheet of polymethylpentene (TPX) attached to the downstream side of
the gate valve. On the other hand, the spin-flipped low-field seekers
remained trapped in the volume between the 6~T field and the UCN
shutter. Opening the shutter allowed the trapped spin-flipped UCNs to
be rapidly drained from the guide system into the detector.

If the UCN gate valve initially remained open for loading time $T_{\rm
  L}$ , followed by cleaning time $T_{\rm C}$ during which time both
the UCN gate valve and the UCN shutter remained closed and at the end
of which the UCN shutter was opened, the number of the detected
spin-flipped neutrons is given by integrating Eq.~(\ref{eq:diff2}):
\begin{equation}
\label{eq:full}
N_{\rm dep} = \beta 
\frac{A_{\rm tot}}{A_{\rm h}}
\int_0^{T_{\rm L}+T_{\rm C}}\int_0^{v_{\rm c}}
P(v,t)R_{\rm h}(t)e^{-\frac{T_{\rm L}+T_{\rm C}-t}{\tau_{\rm dep}(v)}}dv\,dt,
\end{equation}
where $N_{\rm dep} = n_{\rm dep}(t)$ at $t=T_{\rm L}+T_{\rm C}$. Since
both the high-field seekers and the spin-flipped low-field seekers are
detected by the same UCN detector, the possible finite detection
efficiency cancels in Eqs.~(\ref{eq:diff}), (\ref{eq:diff2}), and
(\ref{eq:full}). Measuring $N_{\rm dep}$ and $R_{\rm h}(t)$, with
sufficient knowledge of $P(v,t)$ and $\tau_{\rm dep}(v)$, determines
$\beta$.

The currents for the coils providing the holding field were adjusted
so that the field was higher than 2~mT and the gradient was smaller
than 10~mT/cm to ensure that the probability of Majorana spin flip due
to UCN passing an inhomogeneous field region or due to wall collision
in an inhomogeneous field was less than $10^{-6}$ per
pass or per bounce~\cite{VLA61}.

As seen in Fig.~\ref{fig:schematic2} and
Table~\ref{tab:configurations}, a large portion of the guide system
downstream of the 6~T field was made of copper. Copper was chosen
because its spin-flip probability was determined to be low by previous
measurements~\cite{SER00,SER03}. The surface area of the test sample
was $\sim$35\% of the total surface area of the guide system, diluting
the sensitivity to the spin flip due to the test sample. This was due
to geometrical constraints in the experimental area.

\subsection{Sample preparation}
Two NiP-coated samples were prepared, one on a 316L stainless steel
tube and the other on an aluminum tube. Both tubes had an ID of
72.9~mm. The 316L stainless steel tube was cut from welded tubing that
had been polished and electropolished to roughness average (Ra) of
10~microinches (0.254~$\mu$m), purchased from Valex (specification
401)~\cite{VAL}. The aluminum tube was cut from unpolished 6061-T6511
aluminum tubing.

The coating was applied using high-phosphorus content electroless
nickel phosphorus plating. Electroless nickel plating is an
autocatalytic process in which the deposition of nickel is brought by
chemical reduction of a nickel salt with a reducing
agent~\cite{BRE46,SHA11}. Electroless NiP deposits exhibit numerous
advantages over those done by other methods~\cite{SHA11}. The
advantages include hardness, corrosion and wear resistance, plating
thickness uniformity, and adjustable magnetic and electrical
properties. Because of these advantages, the electroless plating
method is widely used in industry. The coating of our samples was done
by Chem Processing, Inc.~\cite{CHE}. The phosphorus content was
$10.5\pm0.25$~wt.\% ($18.2\pm0.4$~at.\%). The thickness of the coating
was 50~$\mu$m, the largest thickness provided by the vendor, for both
samples. We made this choice as we were interested in seeing if this
coating material could keep UCNs from seeing the stainless steel's
magnetic surface. (We prepared samples with thinner coating but the
limited beam time available to us did not allow us to make
measurements with them.)  Post-baking at 375$\degree$C for 20~hours
was performed for both samples to increase the hardness. Baking also
has the effect of removing hydrogen from the NiP
coating~\cite{APA99}. The coated surface was then cleaned with the
following procedure:
\begin{enumerate}
\item The samples were submerged in Alconox solution (1 wt.\%) at
  60$^{\circ}$C for 10 hours.
\item The samples were then rinsed with deionized water and then with 
  isopropyl alcohol.
\end{enumerate}

\subsection{Measurement}
\begin{figure}
\centering
\includegraphics[width=3in]{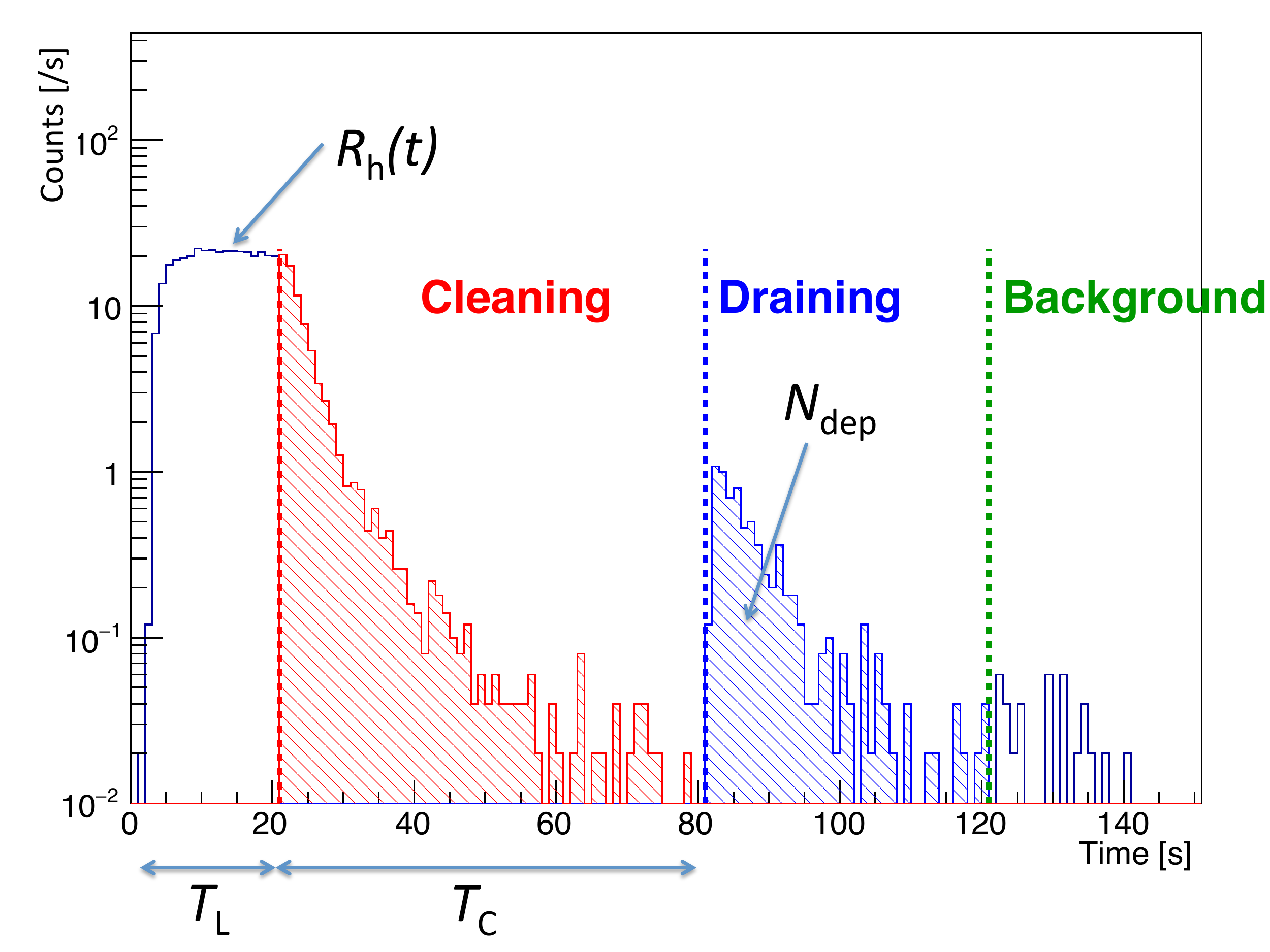}
\caption{Neutron count rate as a function of time obtained for
  configuration C2 for $T_{\rm L} = 20$~s and $T_{\rm C} =
  60$~s. \label{fig:time_spectrum}}
\end{figure}

Measurements were performed for both NiP-coated guide samples. In
addition, to measure the contributions from the rest of the system
including the copper guides and the shutter, measurements were made
with the NiP samples removed. (The limited beam time did not allow us
to perform measurements on uncoated stainless steel samples.) A
description of the three configurations for which measurements were
made is listed in Table~\ref{tab:configurations}. For all
configurations, two series of measurements were performed: one in
which $T_{\rm C}$ was fixed and $T_{\rm L}$ was varied (loading time
scan) and another in which $T_{\rm L}$ was fixed and $T_{\rm C}$ was varied
(cleaning time scan). Table~\ref{tab:TL_TC} lists the values of
$T_{\rm C}$ and $T_{\rm L}$ used in the measurements, and
Fig.~\ref{fig:time_spectrum} shows an example of the measured neutron
count rate as a function of time.

\begin{table*}
\caption{Description of the three configurations for which
  measurements were performed.\label{tab:configurations}}
\begin{tabular}{clcc} \hline\hline
Configuration & Description & {\ }Cu guide length{\ } & {\ }NiP guide
length \\ \hline
C1 & SS guide with 50~$\mu$m NiP coating  & 188 cm & 100 cm \\ 
C2 & Al guide with 50~$\mu$m NiP coating  & 188 cm & 100 cm \\ 
C3 & No NiP-coated guide & 188 cm & 0 cm \\  \hline\hline
\end{tabular}
\end{table*}

\begin{table*}
\caption{The values of $T_{\rm C}$ and $T_{\rm L}$ used for the loading-time and
  the cleaning-time scans.\label{tab:TL_TC}}
\begin{tabular}{lccl} \hline\hline
Scan & $T_{\rm L}$ (s)  & $T_{\rm C}$ (s) \\ \hline
Loading time scan & 10, 20, 40, 60, 80 (C1 only)  & 60 \\ 
Cleaning time scan & 20 (C2), 50 (C1, C3) & 20, 40, 60, 80 \\ \hline\hline
\end{tabular}
\end{table*}

\section{\label{sec:analysis}Data Analysis}
As mentioned earlier, Eq.~(\ref{eq:full}) allows determination of
$\beta$ from experimentally measured $R_{\rm h}(t)$ and $N_{\rm dep}$
if $P(v,t)$ and $\tau_{\rm dep}(v)$ are known with sufficient
accuracy. $P(v,t)$ depends not only on the velocity distribution of
the UCN entering the system but also on the lifetime of high-field
seekers in the system $\tau_{\rm hfs}(v)$. In turn, $\tau_{\rm
  hfs}(v)$, as well as $\tau_{\rm dep}(v)$, depends not only on the
surface-UCN interactions but also on how the system was assembled, as
for many systems the UCN loss is dominated by gaps at joints between
UCN guide sections.

Equation~(\ref{eq:full}) assumes [through Eq.~(\ref{eq:rate})] that
the distribution of UCNs inside the system was uniform and
isotropic. This assumption is justified as follows. Firstly, the $1/e$
time to drain the system was measured to be $\sim$3~s for C1 and C2,
and $\sim$2~s for C3, by fitting the ``Cleaning'' part of the neutron
time spectra (such as the one shown in
Fig.~\ref{fig:time_spectrum}). This indicates that the system reached
a uniform density within 2 to 3~s, a time scale much shorter than the
loading time. Secondly, the transport properties of the UCN guides
that we use are typically well described with Monte Carlo simulations
when we use a nonspecularity of $\epsilon\sim 0.03$ or higher with the
Lambertian angular distributions for nonspecular
reflection~\cite{CLA15}. Since the mean free path between collisions
in a tube of radius $R$ is $2R$, UCNs undergo nonspecular reflection
approximately every $2R/(v\,\epsilon) \sim 0.6$~s. This indicates that
UCNs in the system reached an isotropic distribution in a time period
much smaller than the loading time.

In the absence of sufficiently accurate experimentally obtained
information on $P(v,t)$ and $\tau_{\rm dep}$, we resorted to analysis
models in which approximations were made to Eq.~(\ref{eq:full}) and
evaluated the effect of those approximations to the extracted values
of $\beta$. Below we describe the two specific analysis models we
employed and the results obtained from them.

\subsection{Analysis model 1}
In this model, $\tau_{\rm dep}$ is assumed to be independent of
neutron velocity $v$. With this assumption, Eq.~(\ref{eq:full})
simplifies to
\begin{equation}
\label{eq:simple}
N_{\rm dep} = \beta 
\frac{A_{\rm tot}}{A_{\rm h}}
\int_0^{T_{\rm L}+T_{\rm C}}R_{\rm h}(t)\,e^{-\frac{T_{\rm L}+T_{\rm C}-t}{\tau_{\rm dep}}}dt.
\end{equation}

\begin{figure*}
\centering
\includegraphics[width=3.5in]{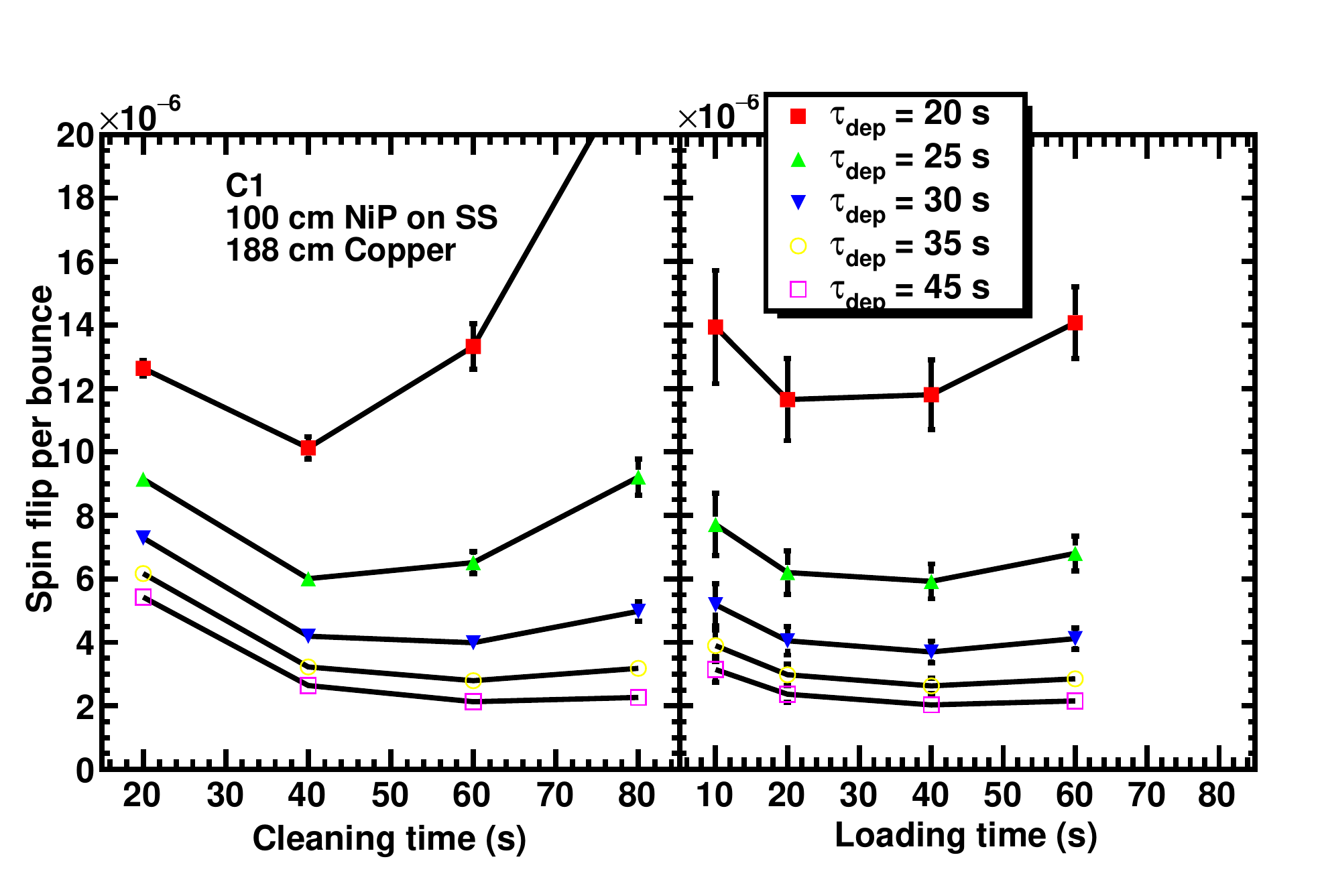}
\includegraphics[width=3.5in]{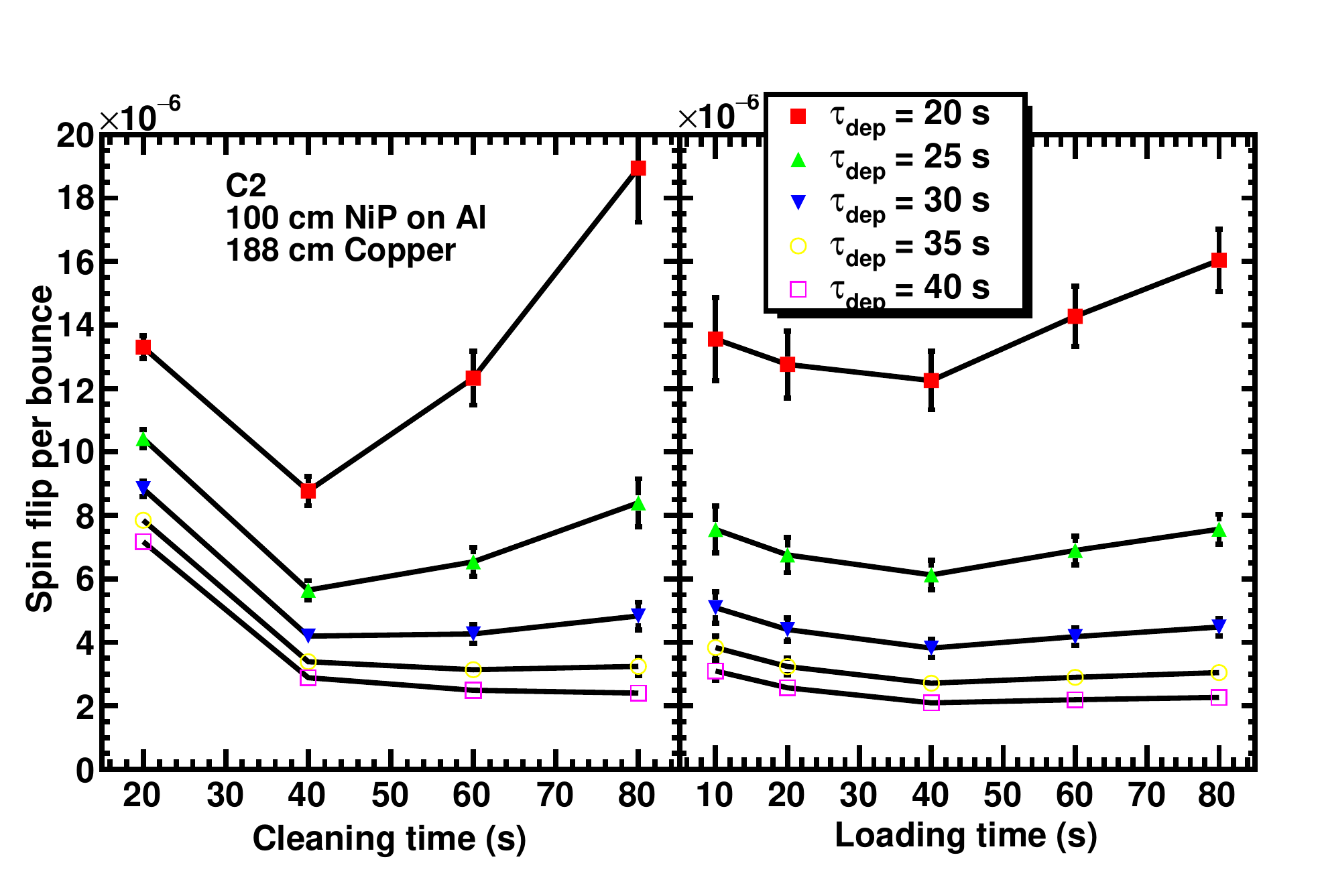}
\includegraphics[width=3.5in]{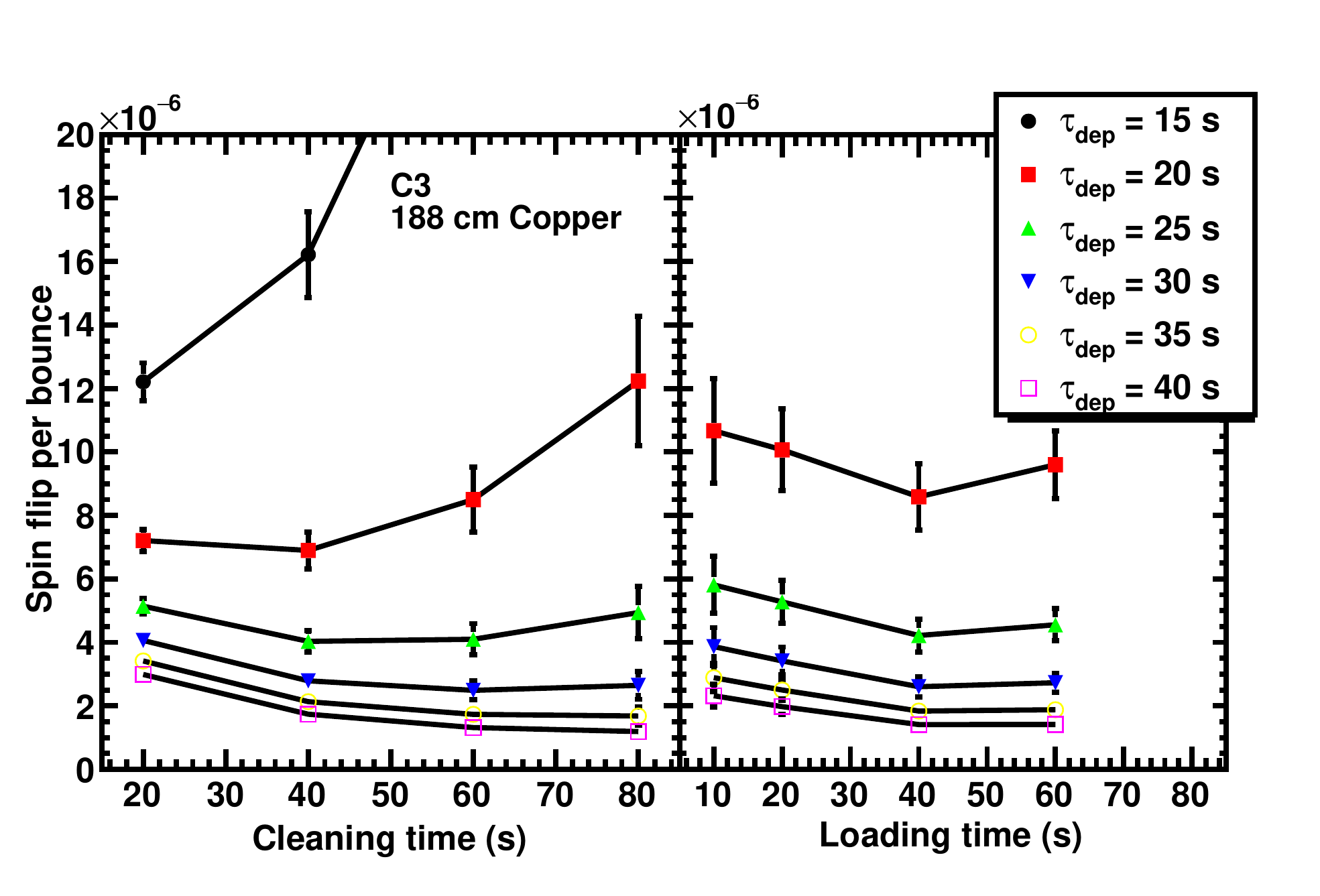}
\caption{Plots of $\beta$ for a set of values of $\tau_{\rm dep}$ for both
  the $T_{\rm L}$ and $T_{\rm C}$ scans obtained assuming
  Eq.~(\ref{eq:simple})\label{fig:simple_results}. The error bar on
  each point represent the statistical uncertainty, which is dominated
  by the counting statistics of $N_{\rm dep}$.}
\end{figure*}

From the experimentally measured $R_{\rm h}(t)$ and $N_{\rm dep}$, we
obtained the values of $\beta$ for a set of assumed values for
$\tau_{\rm dep}$. For the correct value of $\tau_{\rm dep}$, $\beta$
should be independent of $T_{\rm L}$ and $T_{\rm
  C}$. Figure~\ref{fig:simple_results} shows plots of $\beta$ for a
set of values of $\tau_{\rm dep}$ for both the $T_{\rm L}$ and $T_{\rm
  C}$ scans.
In this analysis, $N_{\rm dep}$ was determined by counting UCNs over a
period of 40~s after $t=T_{\rm L}+T_c$ and subtracting the background
estimated from the counts after the counting period.

From these results we see the following:
\begin{itemize}
\item $\beta\sim 4\times 10^{-6}$ can describe data for all
  combinations of $T_{\rm L}$ and $T_{\rm C}$ for all three configurations 
  reasonably well except for $T_{\rm C} = 20$~s. 
\item The deviation for $T_{\rm C}=20$~s is larger for configurations C1 and
  C2 than for configuration C3. 
\end{itemize}
The deviation of the data points for $T_{\rm C}=20$~s from other data points
can be attributed to the fact that 20~s was not sufficiently long to
remove all the high-field seekers, and as a result there were
high-field seekers included in what was counted as $N_{\rm dep}$. That
$T_{\rm C}$ of 20~s was not sufficient to remove all the high-field seekers
can be clearly seen in Fig.~\ref{fig:time_spectrum}.

\begin{figure}
\centering
\includegraphics[width=3in]{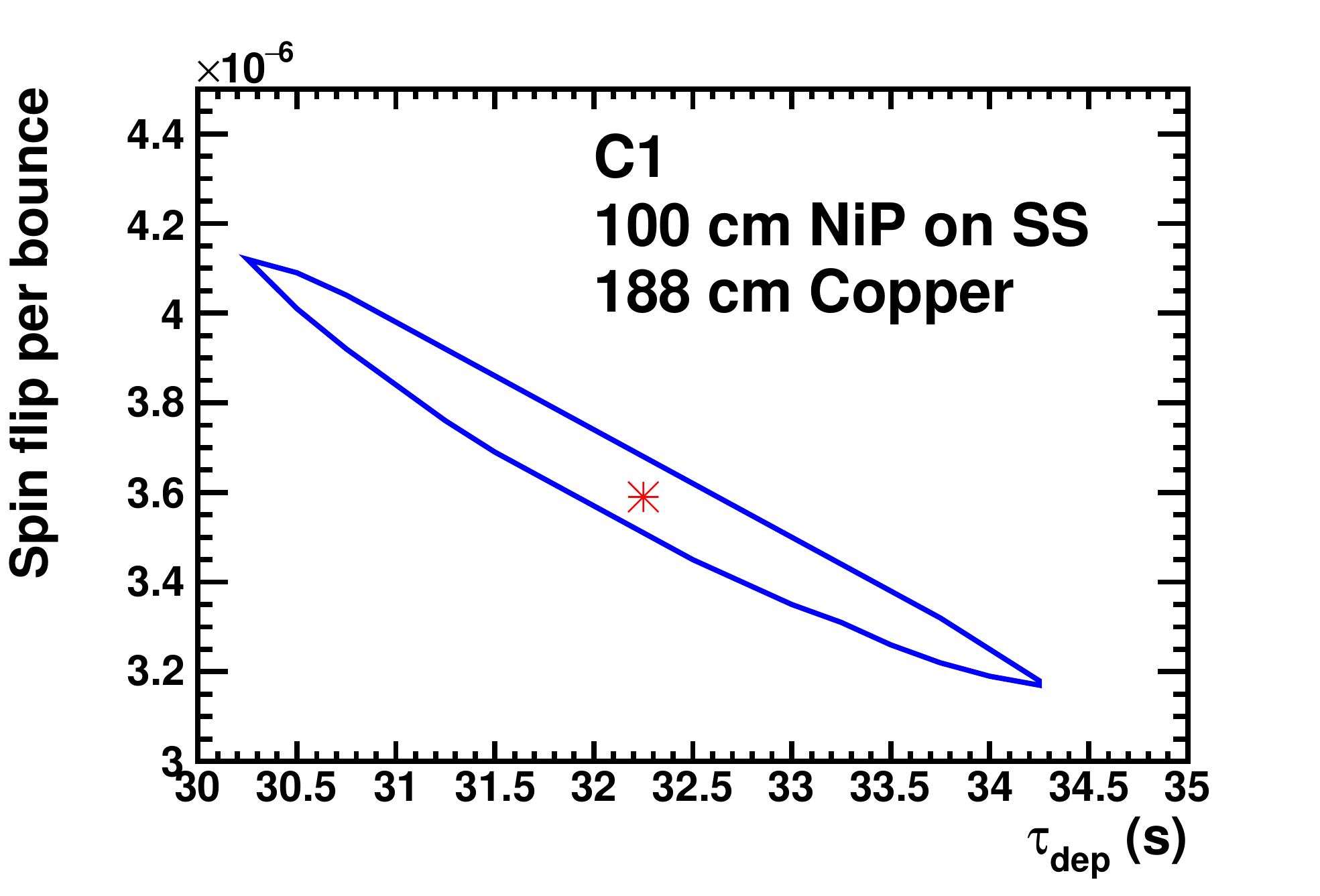}
\includegraphics[width=3in]{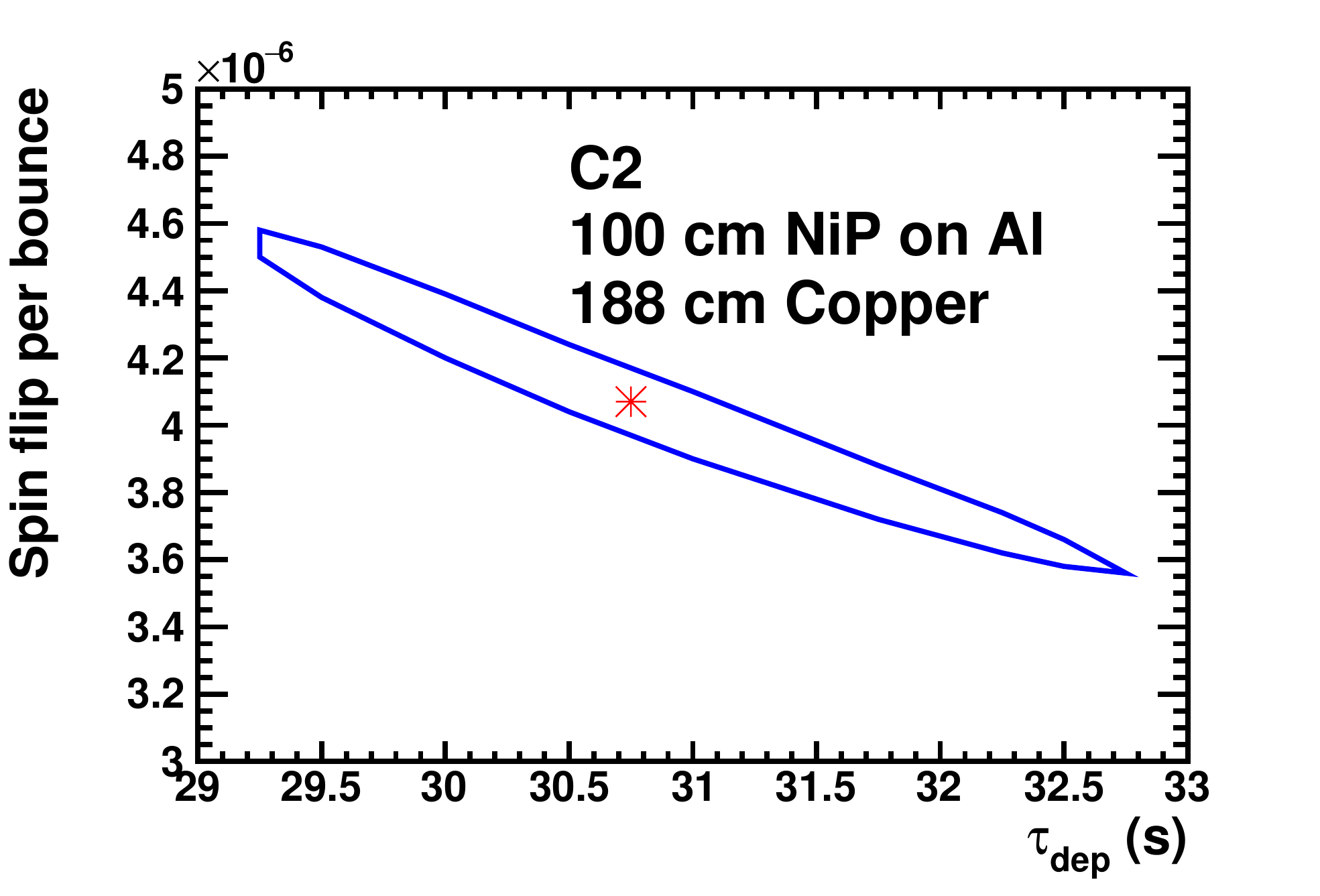}
\includegraphics[width=3in]{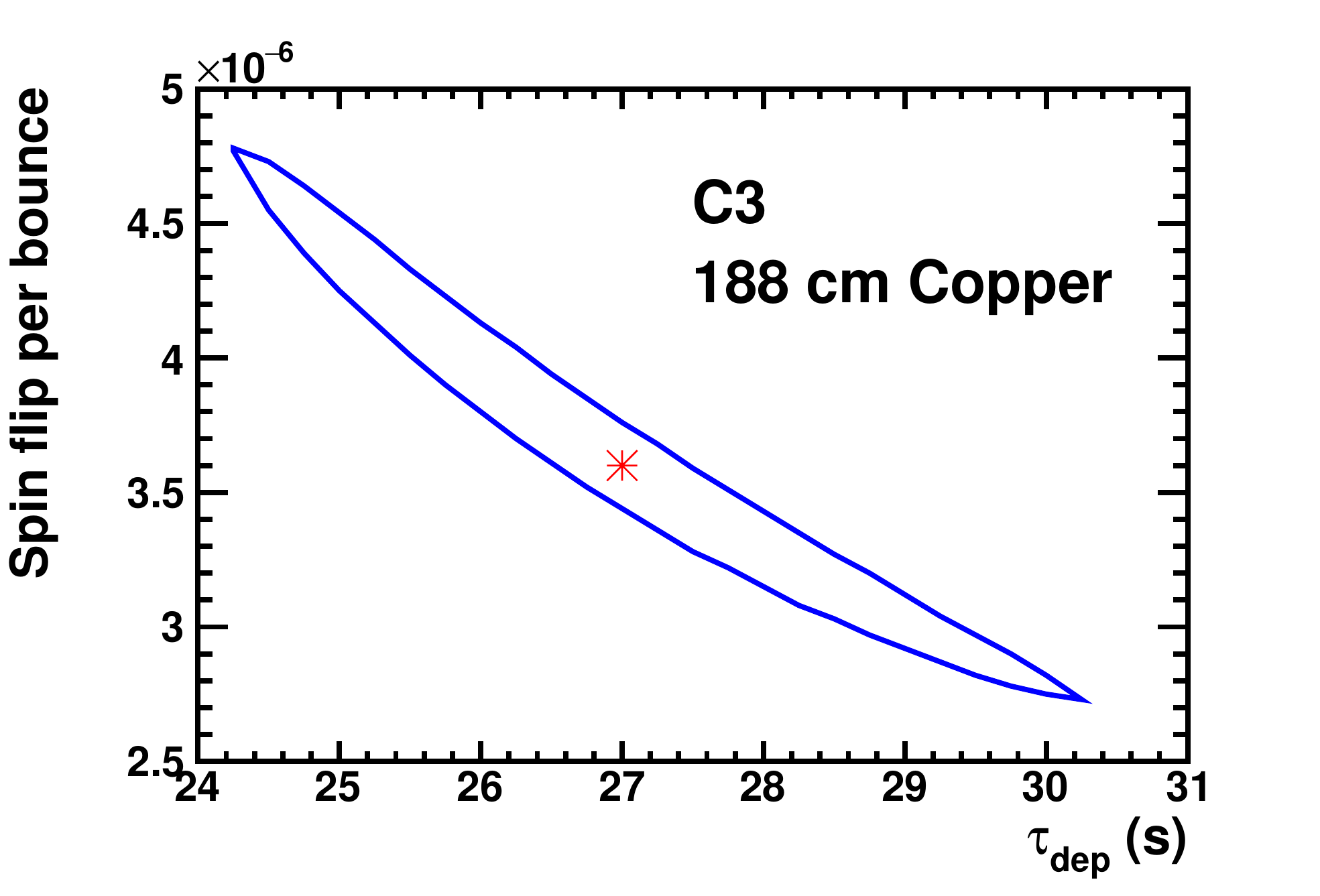}
\caption{best fit values of $\beta$ and $\tau_{\rm dep}$ as well as the 1
  $\sigma$ boundary for each configuration for analysis model
  1.\label{fig:chisq_plus1}}
\end{figure}

The ``best fit'' values of $\beta$ and $\tau_{\rm dep}$ can be obtained by
minimizing $\chi^2$ defined as
\begin{equation}
\chi^2 = \sum_{i=\{T_{\rm C}, T_{\rm L}\}}\frac{\{\beta_i(\tau_{\rm dep})-\beta\}^2}{(\delta\beta_i)^2},
\end{equation}
where the summation is over all combinations of $\{T_{\rm C},T_{\rm
  L}\}$ except for $T_{\rm C}=20$~s. The obtained best fit values of
$\beta$ and $\tau_{\rm dep}$ as well as the 1 $\sigma$ boundary are
graphically indicated in Fig.~\ref{fig:chisq_plus1} for each
configuration. The numerical results are listed in
Table~\ref{tab:AM1}.
\begin{table}
\caption{\label{tab:AM1} Results obtained from Analysis Model 1. Only
  the best fit values are listed for $\tau_{\rm dep}$. For
  description of the guide configurations, see Table~\ref{tab:configurations}.}
\begin{tabular}{lccc}\hline\hline
Guide & $\beta$ & $\tau_{\rm dep}$ (s) & $\chi^2$/DOF\\ \hline
C1 & $(3.6^{+0.5}_{-0.4}) \times 10^{-6}$ & 32.3 & 1.49 \\ 
C2 & $(4.1^{+0.5}_{-0.5}) \times 10^{-6}$ & 30.3& 0.94\\ 
C3 & $(3.6^{+1.2}_{-0.9}) \times 10^{-6}$ & 27.0& 0.84\\ \hline\hline
\end{tabular}
\end{table}

\subsection{Analysis model 2}
In this model, we assumed $P(v)\propto v^2$. The $v^2$ dependence is
well motivated from Monte Carlo simulations performed on the UCN
production and transport~\cite{CLA15}. For the UCN loss, we assumed
\begin{equation}
\tau_{\rm dep}(v) = \left (
\frac{v}{4}\frac{A_{\rm tot}}{V}f
\right )^{-1},
\label{eq:ucn_loss}
\end{equation}
where $V$ is the volume of the trap and $f$ is a factor corresponding
to the probability of UCN loss per collision. In this model we assumed
$f$ to be independent of the UCN energy and the angle of
incidence. This is a good approximation when UCN loss is dominated by
gaps in the system, and not by the interaction of UCNs with walls, as
is the case for this system as discussed in
Sec.~\ref{subsec:discussion}.

Substituting Eq.~(\ref{eq:ucn_loss}) into Eq.~(\ref{eq:full}) with a
model $P(v)=\frac{3}{v_{\rm c}^3}v^2$ and integrating over $v$ yields
\begin{eqnarray}
\label{eq:AM2}
N_{\rm dep} & = & \beta \frac{A_{\rm tot}}{A_{\rm h}}\frac{3}{v_{\rm c}^3}\int_0^{T_{\rm L}+T_{\rm C}}dt
\,R_{\rm h}(t) \nonumber \\
&\times  & \frac{2+e^{-a\,v_{\rm c}}\left \{-2-a\,v_{\rm c} \left (2+a\,v_{\rm c}\right)\right \}}{a^3},
\end{eqnarray}
where $a=\frac{1}{4}\frac{A_{\rm tot}}{V}f\,(T_{\rm L}+T_{\rm C}-t)$.

\begin{figure}
\centering
\includegraphics[width=3in]{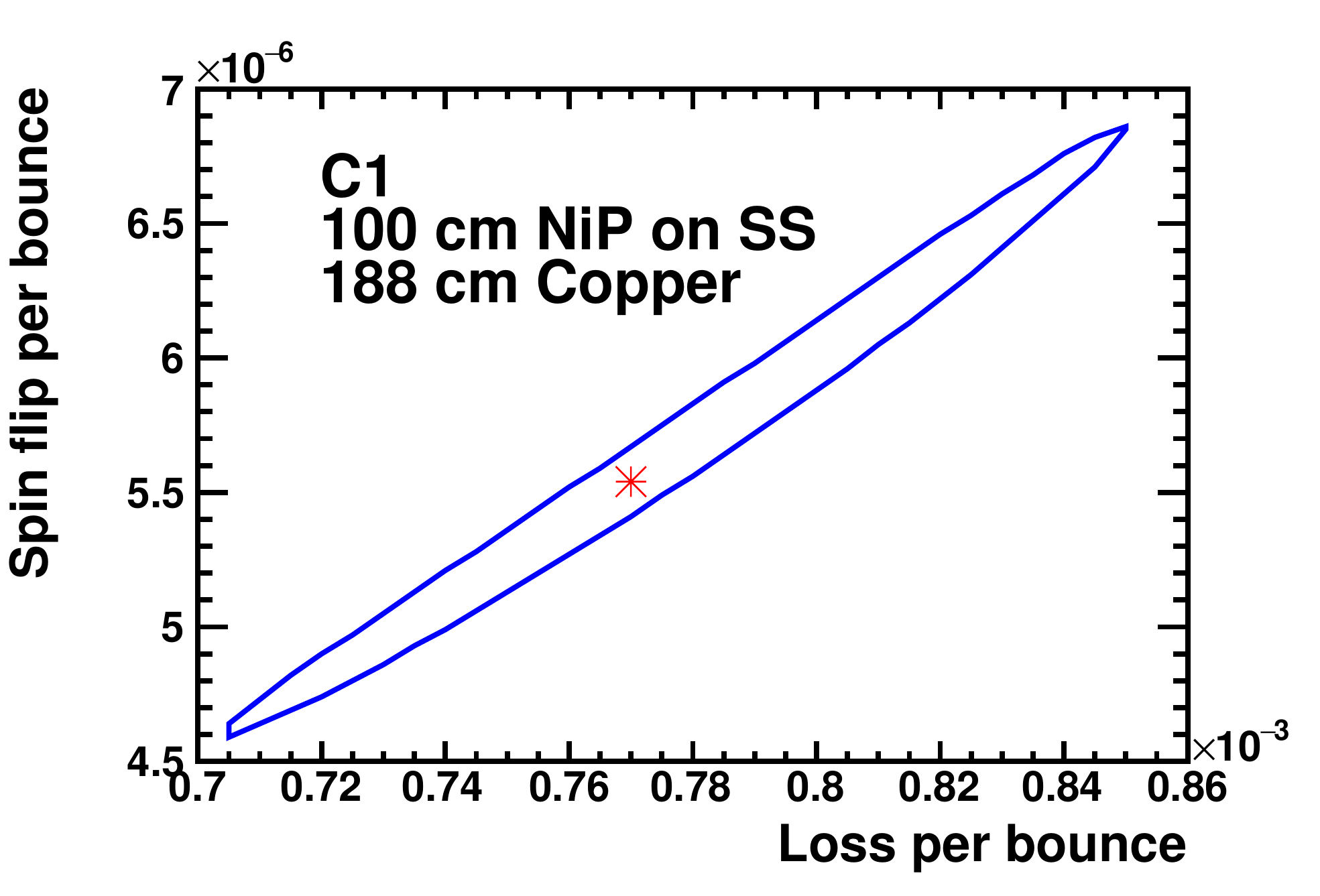}
\includegraphics[width=3in]{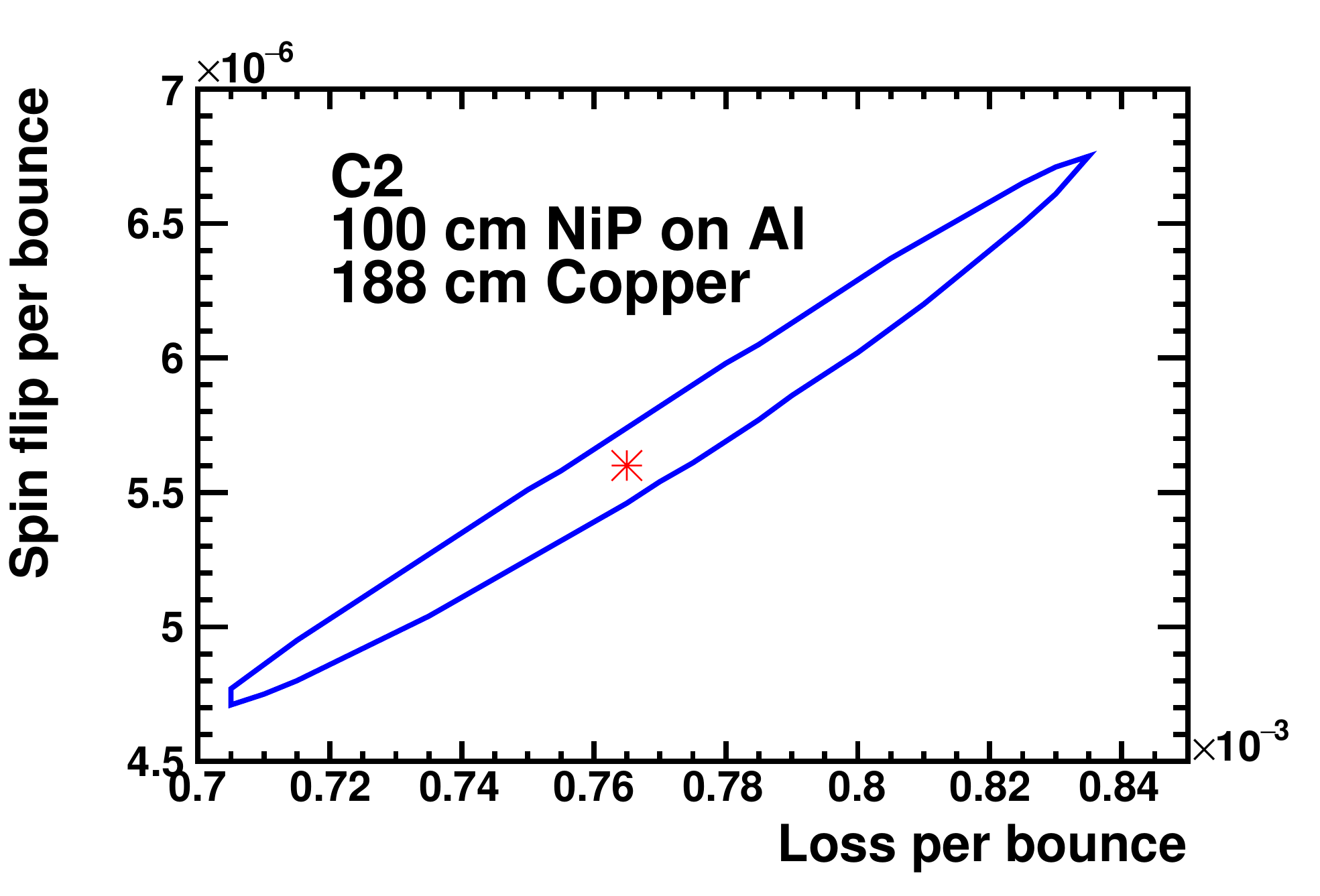}
\includegraphics[width=3in]{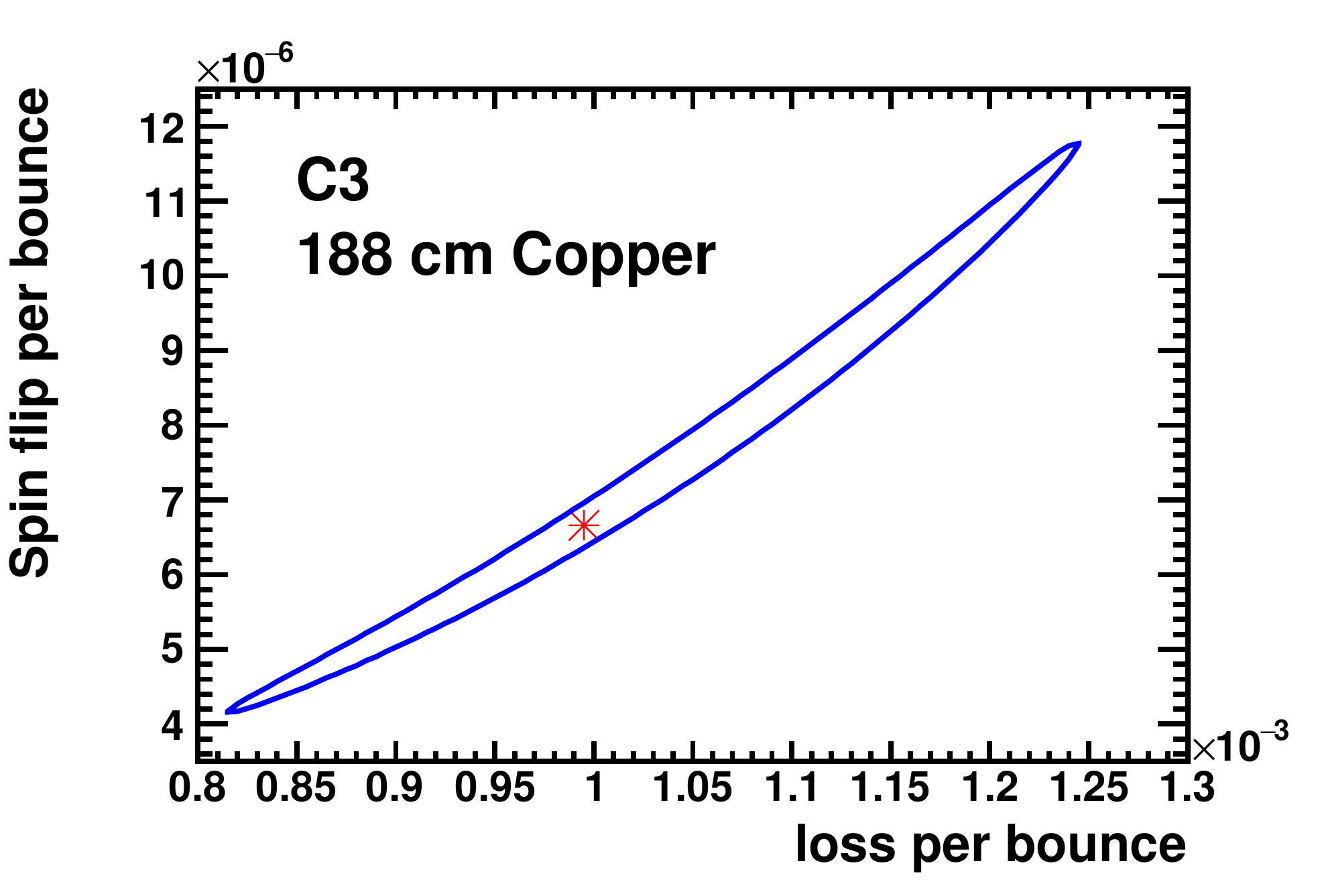}
\caption{best fit values of $\beta$ and $\tau_{\rm dep}$ as well as the 1
  $\sigma$ boundary for each configuration for analysis model
  2.\label{fig:chisq_plus1_3}}
\end{figure}

From the experimentally measured $R_{\rm h}(t)$ and $N_{\rm dep}$, we obtained
the values $\beta$ for a set of assumed values of $f$ using
Eq.~(\ref{eq:AM2}) with $v_{\rm c} = 5.66$~m/s corresponding to the cutoff
energy for copper. The ``best fit'' values of $\beta$ and
$\tau_{\rm dep}$ were obtained by minimizing $\chi^2$ defined as
\begin{equation}
\chi^2 = \sum_{i=\{T_{\rm C}, T_{\rm L}\}}\frac{\{\beta_i(f)-\beta\}^2}{(\delta\beta_i)^2},
\end{equation}
where the summation is over all combinations of $\{T_{\rm C},T_{\rm L}\}$ except
for $T_{\rm C}=20$~s. The obtained best fit values of $\beta$ and $f$ as
well as the 1 $\sigma$ boundary are graphically indicated in
Fig.~\ref{fig:chisq_plus1_3} for each configuration. The numerical
results are listed in Table~\ref{tab:AM2}.
\begin{table}
\caption{\label{tab:AM2} Results obtained from Analysis Model 2. Only
  the best fit values are listed for $f$. For
  description of the guide configurations, see
  Table~\ref{tab:configurations}.}
\begin{tabular}{lccc}\hline\hline
Guide & $\beta$ & $f$ & $\chi^2$/DOF\\ \hline
C1 & $(5.5^{+1.3}_{-1.0}) \times 10^{-6}$ & $7.7\times 10^{-4}$ & 0.99 \\ 
C2 & $(5.6^{+1.2}_{-0.9})\times 10^{-6}$ & $7.7\times 10^{-4}$ & 0.97\\ 
C3 & $(6.7^{+5.0}_{-2.5}) \times 10^{-6}$ & $1.0\times 10^{-3}$ & 0.99\\ \hline\hline
\end{tabular}
\end{table}

\subsection{\label{subsec:discussion}Discussion of the analysis results}
While the two analysis methods yielded somewhat different values for
$\beta$, in each analysis, the difference in $\beta$ among different
guide configurations were small, indicating that the NiP-coated guide
sections were as non-depolarizing as copper guides.

Now we discuss possible systematic biases introduced in the results
due to the simplifying assumptions made for each of the analysis
models.  In Analysis Model 1, it was assumed that the lifetime of the
depolarized neutrons was the same regardless of the UCN velocity. It
is usually the case that slower neutrons have a longer lifetime,
because the loss rate is proportional to the rate of collision with
the walls and also the loss per bounce is larger for faster
neutrons. This effect causes the average velocity of UCNs trapped for
an extended period of time to be lower than the initial average
velocity. Thus the assumption of velocity independent UCN lifetime
ignores the possible spectral difference between the high-field
seekers that initially occupied the system and the detected
spin-flipped UCNs. Since the detected spin-flipped UCNs tended to have
lower velocities than the initial high-field seekers, the estimation
of the collision rate using the measured $R_{\rm h}$ is an
overestimate for the detected spin-flipped UCNs. This results in an
underestimate of the spin flip probability $\beta$.

Analysis Model 2 addresses this problem. It describes UCN loss with a
single parameter, the loss per bounce $f$, which is independent of the
UCN velocity. As mentioned earlier, this is a good approximation when
the UCN loss is dominated by gaps in the system. In fact, the obtained
values of $f$ are $\sim$10$^{-3}$, much larger than the typical loss
per bounce due to collision with material walls. Here, $f\sim 10^{-3}$
indicates a 1~mm wide gap for every 1~m of guide, which agrees with
previous assessments of gaps in typical guide assemblies of this
type. The contribution to $f$ from the pinhole is much smaller,
$\approx 2.9 \times 10^{-5}$ for C1 and C2 and $\approx 1.3 \times
10^{-5}$ for C3.

The values of $\beta$ obtained from Analysis Model 2 are larger than
those from Analysis 1 by a factor of 1.4-1.9. This can be explained by
the underestimation of $\beta$ in Analysis Model 1 that was discussed
earlier. The size of this effect can be estimated as follows. For a
neutron velocity spectrum proportional to $v^2$, the average velocity
is $v_{\rm ave} = \frac{3}{4}v_{\rm c}$. If the UCN loss is described
by a constant loss per bounce $f$, then the average velocity of the
UCN after time $t$ is
\begin{eqnarray}
v_{\rm ave}(t) & = & \frac{\int_0^{v_{\rm c}} v^3 e^{-b\,v\,t}\,dv}{\int_0^{v_{\rm c}} v^2 e^{-b\,v\,t}\,dv}
\nonumber \\
& = & 
\frac{6+e^{-c}[-6-c\,\{6+c\,(3+c)\}]}{b\,t\,[2+e^{-c}\{-2-c\,(2+c)\}]},
\end{eqnarray}
where $b=\frac{1}{4}\frac{A_{\rm tot}}{V}f$ and $c=b\,t\,v_{\rm
  c}$. Inserting the values from this experiment of $v_{\rm
  c}=5.66$~m/s, $4V/A_{\rm tot}=7.62$~cm, and $f=0.7\times 10^{-3}$
gives $\frac{v_{\rm ave}(t)}{v_{\rm ave}(0)} \simeq 1.5$ for {\it
  e.g.} $t=100$~s, consistent with the observation made with the
comparison between Analysis Models 1 and 2.

Both analysis results indicate that the UCN loss was smaller when a
NiP-coated guide section was included in the system than it was when
the system was solely made of copper guides. This can be attributed to
two factors: (1)~the smaller average number of gaps per unit length
when the NiP-coated guide is included in the system as the NiP-coated
section is longer than the other guide section (see
Fig.~\ref{fig:schematic2}), and (2)~the superior UCN storage
properties of NiP-coated guides that we observed~\cite{PAT15}, which
is due to the higher Fermi potential of NiP.

Assuming that UCNs were uniformly distributed inside the system we can
relate the spin flip probabilities obtained for the three
configurations to separate contributions from NiP-coated
and copper surfaces:
\begin{equation}
\beta_{\rm C1,C2} =  f_{\rm Cu}\, \beta_{\rm Cu} + f_{\rm NiP}
\,\beta_{\rm NiP},
\label{eq:beta}
\end{equation}
where $f_{\rm Cu}$ and $f_{\rm NiP}$ are the fraction of the copper
surface and that of the NiP-coated surface in the system with $f_{\rm
  NiP} = 0.35$ and $f_{\rm Cu} = 0.65$ for configurations C1 and C2,
and $\beta_{\rm Cu} = \beta_{\rm C3}$. The results obtained from
Eq.~(\ref{eq:beta}) and Analysis Model 2 (Table~\ref{tab:AM2}) are
shown in Table~\ref{tab:results}. Here we assumed that the
uncertainties on $\beta_{\rm C1}$, $\beta_{\rm C2}$, and $\beta_{\rm
  C3}$, which are dominated by the uncertainties on the loss per
bounce parameters, are 100\% correlated.

The resulting $\beta$ for the NiP-coated guides include the unphysical
region $\beta < 0$ within 1 $\sigma$. They can be interpreted as upper
limits, giving $\beta_{\rm NiP\;on\;SS} < 6.2 \times 10^{-6}$ (90\%
C.L.) and $\beta_{\rm NiP\;on\;Al} < 7.0 \times 10^{-6}$ (90\%
C.L.). Here we followed the Feldman-Cousins prescription~\cite{FEL98}.

\begin{table}
\caption{\label{tab:results} Values of the spin-flip probability per
  bounce $\beta$ determined from our work}
\begin{tabular}{lc}\hline\hline
Material & $\beta$ \\ \hline
NiP on 316L SS & $(3.3^{+1.8}_{-5.6}) \times 10^{-6}$\\ 
NiP on Al & $(3.6^{+2.1}_{-5.9}) \times 10^{-6}$ \\ 
Copper & $(6.7^{+5.0}_{-2.5}) \times 10^{-6}$ \\ \hline\hline
\end{tabular}
\end{table}

The uncertainties of our results were dominated by the uncertainty on
the value of the loss per bounce for UCN stored in the system, in
particular that for C3, the contribution of which had to be subtracted
from the results obtained for C1 and C2. For future experiments, the
uncertainty can be significantly reduced by making the entire guide
system out of NiP-coated guide sections. The uncertainty can be
further reduced by making an auxiliary measurement to determine the
loss per bounce parameter. This could be achieved by, for example,
performing a measurement without opening the shutter and observing the
rate at which the depolarized neutrons leak through the pinhole on the
shutter, if there is a sufficiently high UCN density stored in the
system. With such an auxiliary measurement, it would be possible to
determine the value of $\beta$ on NiP-coated surfaces to better than
10\%.

In addition, studying the energy dependence of the depolarization
probability is of great interest, as it may shed light on the
mechanism of UCN depolarization. Such a study can be performed by
repeating measurements described in this paper for different field
strength settings for the superconducting magnet, thus varying the
cutoff energy of the stored neutrons as was done in
Ref.~\cite{ATC07}. Such a study of energy dependence would also allow
a determination of the energy dependence of the UCN loss parameter.

\section{\label{sec:discussion}Discussion}
Our results indicate that the depolarization per bounce on the surface
of NiP-coated 316L SS as well as NiP-coated Al is as small as that of
copper surfaces. 316L stainless steel is austenitic and is considered
to be ``non-magnetic.'' However, stress due to cold working can cause
it to develop martensitic microstructure. Such surface magnetism is
likely to be the reason for the large depolarization probability
observed previously.  The 50~$\mu$m thick NiP coating seems to be
sufficiently thick to keep UCNs from ``seeing'' the magnetic surface.

As mentioned earlier, NiP coating is extremely robust and is widely
used in industrial applications. The coating can be applied
commercially in a rather straightforward chemical process and there
are numerous vendors that can provide such a service
inexpensively. The ability of NiP coatings to turn SS surfaces
non-depolarizing has many practical applications. SS UCN guides are
widely used in UCN facilities because of its mechanical robustness and
relatively high UCN potential. However, because of its highly
depolarizing nature, other guide materials, such as quartz coated with
DLC or nickel molybdenum, have so far been used when non-depolarizing
guides were needed. Simple NiP coatings can dramatically reduce the
depolarization on SS surfaces. In addition, NiP has a higher Fermi
potential than SS. We have confirmed that SS flanges retain their
sealing properties even after they are coated with NiP. Therefore
NiP-coated SS guides can be used where both mechanical robustness as
well as non-depolarizing properties are required.

\section{\label{sec:summary}Summary}
We measured the spin-flip probabilities for ultracold neutrons
interacting with surfaces coated with nickel phosphorus.  For
50~$\mu$m thick nickel phosphorus coated on stainless steel, the
spin-flip probability per bounce was found to be $\beta_{\rm
  NiP\;on\;SS} = (3.3^{+1.8}_{-5.6}) \times 10^{-6}$. For 50~$\mu$m
thick nickel phosphorus coated on aluminum, the spin-flip probability
per bounce was found to be $\beta_{\rm NiP\;on\;Al} =
(3.6^{+2.1}_{-5.9}) \times 10^{-6}$. For the copper guide used as
reference, the spin flip probability per bounce was found to be
$\beta_{\rm Cu} = (6.7^{+5.0}_{-2.5}) \times 10^{-6}$.  The results on
the nickel phosphorus-coated surfaces may be interpreted as upper
limits, yielding $\beta_{\rm NiP\;on\;SS} < 6.2 \times 10^{-6}$ (90\%
C.L.) and $\beta_{\rm NiP\;on\;Al} < 7.0 \times 10^{-6}$ (90\% C.L.)
for 50~$\mu$m thick nickel phosphorus coated on stainless steel and
50~$\mu$m thick nickel phosphorus coated on aluminum, respectively.

Our results indicate that NiP coatings can make SS surfaces
competitive with copper for experiments that require maintaining high
UCN polarization.  Because NiP has a higher Fermi potential than
copper, can be used with SS guide tubes, has excellent corrosion
resistance, and coatings are available commercially, it may become the
"coating of choice" for a number of applications in experiments
conducted with polarized UCNs.

\section*{Acknowledgments}
This work was supported by Los Alamos National Laboratory LDRD Program
(Project No. 20140015DR). We gratefully acknowledge the support
provided by the LANL Physics and AOT Divisions.


\section*{References}
\begin{thebibliography}{00}

\bibitem{DUB11}
D.~Dubbers and M.~G.~Schmidt, Rev. Mod. Phys. {\bf 83}, 1111 (2011).

\bibitem{YOU14}
A.~R.~Young, {\it et al.} J. Phys. G: Nucl. Part. Phys. {\bf 41},
114007 (2014).

\bibitem{STY86}
A.~Steyerl, {\it et al.}, Phys. Lett. A {\bf 116}, 347 (1986)

\bibitem{SAU13}
A.~Saunders, {\it et al.}, Rev. Sci. Instrum. {\bf 84}, 013304 (2013).

\bibitem{MAS12}
Y.~Masuda, {\it et al.}, Phys. Rev. Lett. {\bf 108}, 134801 (2012)

\bibitem{BEC15}
H.~Becker, {\it et al.} Nucl. Instrum. Methods Phys. Res. A {\bf 777},
20 (2015).

\bibitem{LAU13}
Th.~Lauer and Th.~Zechlau, Eur. Phys. J. A {\bf 49}, 104 (2013).

\bibitem{SAL14}
D.~J.~Salvat, {\it et al.}, Phys. Rev. C {\bf 89}, 052501(R) (2014).

\bibitem{BAK06}
C.~A.~Baker, {\it et al.}, Phys. Rev. Lett {\bf 97}, 131801 (2006).

\bibitem{PLA12}
B.~Plaster, {\it et al.}, Phys. Rev. C {\bf 86}, 055501 (2012).

\bibitem{MEN13}
M.~Mendenhall, {\it et al.}, Phys. Rev. C {\bf 87}, 032501(R) (2013).

\bibitem{SER00}
A.~Serebrov, {\it et al.}, Nucl. Instrum. Methods Phys. Res. A {\bf
  440}, 717 (2000).

\bibitem{POK02}
Yu.~N.~Pokotilovski, JETP Lett. {\bf 76}, 162 (2002).

\bibitem{SER03}
A.~P.~Serebrov, {\it et al.} Phys. Lett. A {\bf 313}, 373 (2003).

\bibitem{ATC07}
F.~Atchison, {\it et al.} Phys. Rev. C {\bf 76}, 044001 (2007).

\bibitem{MAK05}
M.~Makela, PhD Thesis, Virginia Polytechnic Institute and State
University, 2005.

\bibitem{RIO09}
R.~Rios, APS Meeting Abstract BAPS.2009.APS.C14.3 (2009).

\bibitem{HOL12}
A.~T.~Holley, PhD Thesis, North Carolina State University (2012).

\bibitem{VLA61}
V.~V.~Vladimirskii, Sov. Phys. JETP {\bf 12}, 740 (1961).

\bibitem{GAM65}
R.~L.~Gamblin and T.~R.~Carver, Phys. Rev. {\bf 138} A946 (1965).

\bibitem{MAS13}
Y.~Masuda (private communication).

\bibitem{PAT15}
R.~W.~Pattie Jr., {\it et al.}, in preparation. 

\bibitem{ALB67}
P.~A.~Albert, Z.~Kovac, H.~R.~Lilenthal, T.~R.~McGuire, and
Y.~Nakamura, J. Appl. Phys. {\bf 38}, 1258 (1967).

\bibitem{BER78}
A.~Berrada, M.~F.~Lapierre, B.~Loegel, P.~Panissod, and C.~Robert,
J. Phys. F {\bf 8}, 845 (1978).

\bibitem{HUM98}
E.~Humbert, A.~J.~Tosser, J. Mat. Sci. Lett. {\bf 17}, 167 (1998).

\bibitem{ITO14} Los Alamos National Laboratory LDRD Project
  \#20140015DR, ``Probing New Sources of Time-Reversal Violation with
  Neutron EDM'', Takeyasu Ito, PI.

\bibitem{ITO15}
T.~M.~Ito, {\it et al.}, in preparation.

\bibitem{WAN15}
Z.~Wang, {\it et al.} Nucl. Instrum. Methods Phys. Res. Sect. A {\bf
  798}, 30 (2015).

\bibitem{VAL}
Valex Corp. {\tt http://www.valex.com}.

\bibitem{BRE46}
A.~Brenner and G.~E.~Riddell, J. Res. Nat. Bur. Stand., {\bf 37}, 91
(1946).

\bibitem{SHA11}
W.~Sha, X.~Wu, and K.~G.~Keong, {\it Electroless Copper and
  Nickel-Phosphorus Plating} (Woodhead Publishing, Cambridge, United
Kingdom, 2011).

\bibitem{CHE}
Chem Processing, Inc. {\tt http://www.chemprocessing.com}.

\bibitem{APA99}
I.~Apachitei and J.~Duszczyk, J. Appl. Electrochem. {\bf 29}, 837 (1999).

\bibitem{CLA15}
S.~M.~Clayton (unpublished).

\bibitem{FEL98}
G.~J.~Feldman and R.~D.~Cousins, Phys. Rev. D {\bf 57}, 3873 (1998).


\end{thebibliography}
\end{document}